\documentclass[aps,prl,twocolumn,superscriptaddress]{revtex4-2}

\usepackage{multirow}
\usepackage{graphicx}
\usepackage{bm} 
\usepackage{amsmath,amssymb}
\usepackage{hyperref}
\usepackage{lipsum}
\usepackage{color}

\usepackage{bm}
\usepackage{textcase}
\usepackage{textcomp}
\usepackage{braket}
\usepackage{setspace}
\usepackage{mathtools}
\usepackage[markup=default]{changes}

\newcommand{\SIQSE}{\affiliation{1}{Shenzhen Institute for Quantum Science and Engineering, Southern University of Science and Technology, Shenzhen 518055, China}}
\newcommand{\IQA}{\affiliation{2}{International Quantum Academy, Shenzhen 518048, China}}
\newcommand{\GDKL}{\affiliation{3}{Guangdong Provincial Key Laboratory of Quantum Science and Engineering, Southern University of Science and Technology, Shenzhen 518055, China}}
\newcommand{\NXU}{\affiliation{4}{
School of Physics, Ningxia University, Yinchuan 750021, PR China}}

\newcommand{\HFNL}{\affiliation{6}{
Shenzhen Branch, Hefei National Laboratory, Shenzhen 518048, China}}

\begin{document}
\title{High-performance multiplexed readout of superconducting qubits with a tunable broadband Purcell filter}

\author{Yuzhe Xiong}
\thanks{These authors contributed equally to this work.}
\affiliation{\SIQSE}\affiliation{\IQA}\affiliation{\GDKL}
\author{Zilin Wang}
\thanks{These authors contributed equally to this work.}
\affiliation{\NXU}\affiliation{\IQA}

\author{Jiawei Zhang}
\affiliation{\SIQSE}\affiliation{\IQA}\affiliation{\GDKL}

\author{Xuandong Sun}
\affiliation{\SIQSE}\affiliation{\IQA}\affiliation{\GDKL}

\author{Zihao Zhang}
\affiliation{\SIQSE}\affiliation{\IQA}\affiliation{\GDKL}

\author{Peisheng Huang}
\affiliation{\NXU}\affiliation{\IQA}

\author{Yongqi Liang}
\affiliation{\SIQSE}\affiliation{\IQA}\affiliation{\GDKL}

\author{Ji Jiang}
\affiliation{\IQA}

\author{Jiawei Qiu}
\affiliation{\IQA}

\author{Yuxuan Zhou}
\affiliation{\IQA}

\author{Xiayu Linpeng}
\affiliation{\IQA}

\author{Wenhui Huang}
\affiliation{\IQA}

\author{Jingjing Niu}
\affiliation{\IQA}\affiliation{\HFNL}

\author{Youpeng Zhong}
\email{zhongyoupeng@iqasz.cn}
\affiliation{\SIQSE}\affiliation{\IQA}\affiliation{\NXU}\affiliation{\HFNL}

\author{Ji Chu}
\email{jichu@iqasz.cn}
\affiliation{\IQA}

\author{Song Liu}
\affiliation{\IQA}\affiliation{\HFNL}

\author{Dapeng Yu}
\affiliation{\IQA}\affiliation{\NXU}\affiliation{\HFNL}

\date{\today}

\begin{abstract}

Fast, high-fidelity, and low back-action readout plays a crucial role in the advancement of quantum error correction (QEC). Here, we demonstrate high-performance multiplexed readout of superconducting qubits using a tunable broadband Purcell filter, effectively resolving the fundamental trade-off between measurement speed and photon-noise-induced dephasing. By dynamically tuning the filter parameters, we suppress photon-noise-induced dephasing by a factor of 7 in idle status, while enabling rapid, high-fidelity readout in measurement status. We achieve 99.6\% single-shot readout fidelity with 100~ns readout pulse, limited primarily by relaxation errors during readout. Using a multilevel readout protocol, we further attain 99.9\% fidelity in 50~ns. Simultaneous readout of three qubits using 100~ns pulses achieves an average fidelity of 99.5\% with low crosstalk. Additionally, the readout exhibits high quantum-nondemolition (QND) performance: 99.4\% fidelity over repeated measurements and a low leakage rate below 0.1\%. Building on the tunable broadband filter, we further propose a scalable readout scheme for surface code QEC with enhanced multiplexing capability, offering a promising solution for fast and scalable QEC.

\end{abstract}
\maketitle

Quantum error correction (QEC) is an essential framework for fault-tolerant quantum computation~\cite{knill1997theory,terhal2015quantum,roffe2019quantum,gidney2025factor}. Although substantial experimental advances have been demonstrated across various quantum computing platforms~\cite{egan2021fault,postler2022demonstration,ryan2024high,pogorelov2025experimental,paetznick2024demonstration,bluvstein2024logical,abobeih2022fault,zhao2022realization,krinner2022realizing,google2023suppressing,google2025quantum}, qubit readout performance has emerged as a critical bottleneck. This limitation is particularly pronounced during stabilizer measurements~\cite{fowler2012surface}, where ancilla qubit readout operates more than an order of magnitude slower than quantum gate operations, constraining further enhancements in QEC performance.

In superconducting circuits, dispersive readout serves as the established method for high-fidelity qubit state measurement~\cite{blais2021circuit}. 
The measurement speed can be enhanced through two approaches: increasing the readout power or strengthening the interactions, specifically the qubit-resonator dispersive shift \(\chi\) and the resonators' external coupling rate $\kappa_{\rm r}$~\cite{Gambetta2006,clerk2010introduction}. 
Increasing readout power induces qubit state leakage~\cite{sank2016measurement,khezri2023measurement,hazra2025benchmarking}, which substantially degrades QEC performance~\cite{mcewen2021removing,miao2023overcoming}.
Conversely, strengthening the interaction aggravates decoherence via the readout channel, including Purcell decay~\cite{purcell1995spontaneous,houck2008controlling} and photon-noise-induced dephasing~\cite{bertet2005dephasing,Clerk2007,sears2012photon,yan2016flux,yan2018distinguishing,zhang2017suppression,wang2019cavity}. 
Purcell decay can be effectively mitigated using Purcell filters that suppress photon emission at the qubit frequency~\cite{sete2015quantum,walter2017rapid,jeffrey2014fast}, and shared broadband Purcell filters have gained widespread adoption for their superior scalability~\cite{arute2019quantum,gong2021quantum,google2021exponential,yan2023broadband,zhou2024high}.
The dephasing, caused by residual photons in the resonator from thermal noise or parasitic measurements~\cite{heinsoo2018rapid}, ultimately prevents further enhancement of the dispersive interaction.
Consequently, the fundamental trade-off between measurement speed and photon-noise-induced dephasing continues to be a key constraint on readout performance.
In response to this challenge, tunable readout schemes have recently emerged as a promising solution~\cite{sunada2024photon,wang2025longitudinal,xiao2025flexible}.

\begin{figure}[t]
    \centering
    \includegraphics[width=\columnwidth]{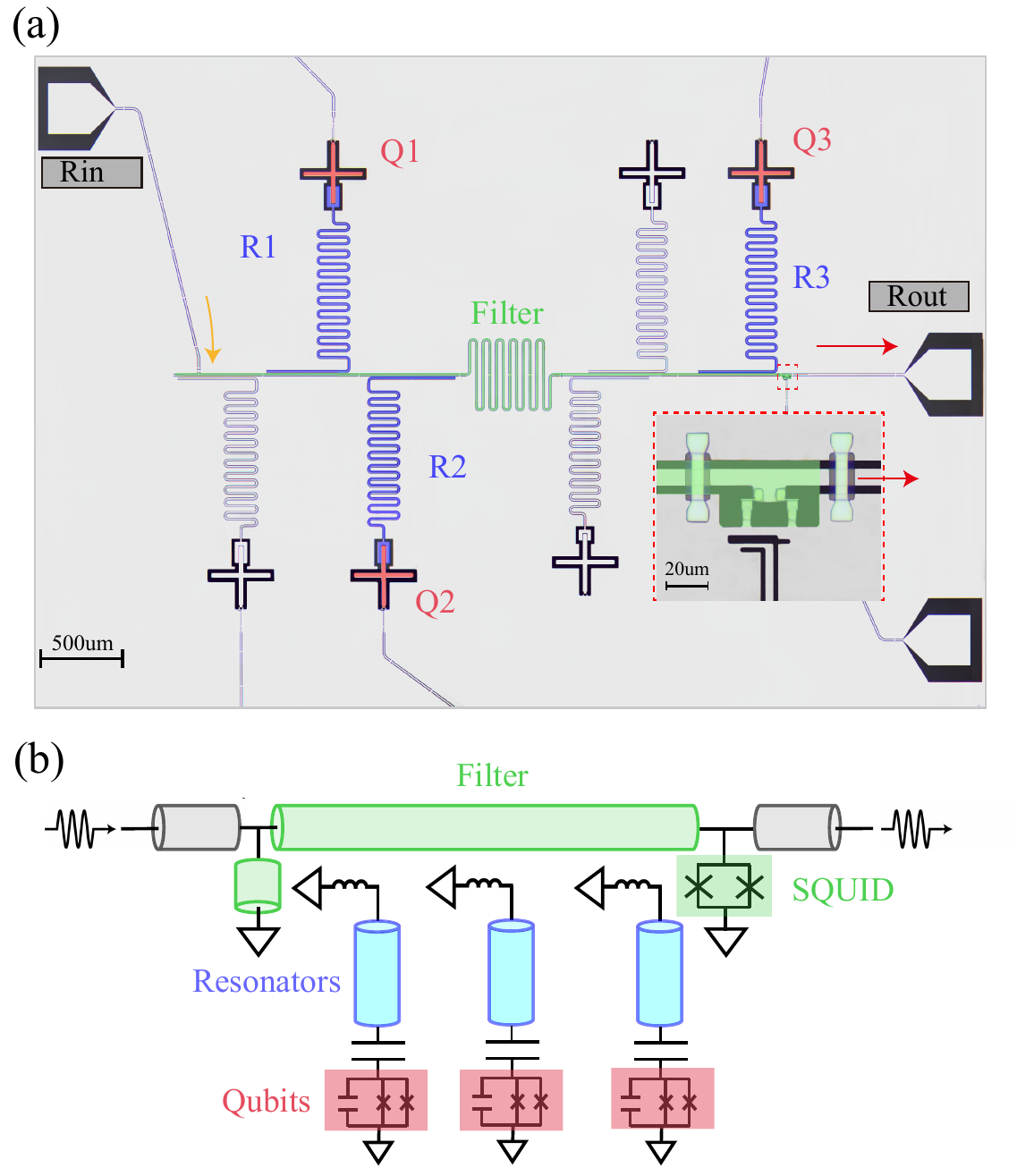}
    \caption{Device layout and schematic.
    (a) False-color optical micrograph of the superconducting chip. Key components are highlighted: the half-wavelength tunable Purcell filter (green), quarter-wavelength readout resonators (blue), and transmon qubits used in this work (red). Unused qubits are left uncolored. The inset shows a zoomed-in view of the SQUID at the filter's shorted end, which is controlled by a dedicated flux line. Arrows indicate the readout signal path (yellow: input, red: output).
    (b) Distributed-element circuit model of the device, with colors corresponding to the physical elements in (a).
     \label{fig1}}
\end{figure}

In this work, we demonstrate high-performance multiplexed readout of superconducting qubits using a tunable broadband Purcell filter. By dynamically controlling the effective linewidth $\kappa_\mathrm{r}$ of all resonators through a shared filter, we overcome the inherent trade-off between readout speed and photon-noise-induced dephasing.
During idling periods, the filter frequency is detuned from the resonators and its bandwidth is narrowed, suppressing the resonator linewidth to $\kappa_{\mathrm{r}} \ll \chi$ and thereby significantly enhancing the qubit dephasing time. 
For rapid high-fidelity readout, the filter is tuned into resonance with the resonator band, increasing $\kappa_\mathrm{r}$ to approximately \(2\chi\). 
This configuration enables 99.6\% fidelity with 100~ns readout pulses, where performance is limited by qubit relaxation during measurement. Using multilevel readout techniques, we further overcome the relaxation limit and achieve 99.9\% fidelity with 50~ns pulses.
The multiplexing capability is demonstrated via simultaneous readout of three qubits using 100~ns pulses, yielding an average fidelity of 99.5\% with minimal crosstalk. 
Furthermore, we introduce a simple leakage benchmarking protocol based on repetitive measurements interleaved with X gates, which reveals a leakage rate below 0.1\% for 100~ns readout---corresponding to a QND fidelity of 99.4\%---and confirms the absence of detrimental back-action from the nonlinear filter.
Based on these results, we further propose a scalable readout architecture for surface-code-based QEC, in which resonators for data and ancilla qubits share a common readout line while operating in disjoint frequency bands. The scheme enhances readout multiplexing capacity while suppressing parasitic measurement-induced dephasing of data qubits during rapid stabilizer measurements. 
Our results establish tunable broadband Purcell filters as a key enabling technology for rapid, high-fidelity, low-back-action readout, highlighting their crucial potential for large-scale quantum error correction.

\begin{figure*}[t]
    \centering
    \includegraphics[width=\textwidth]{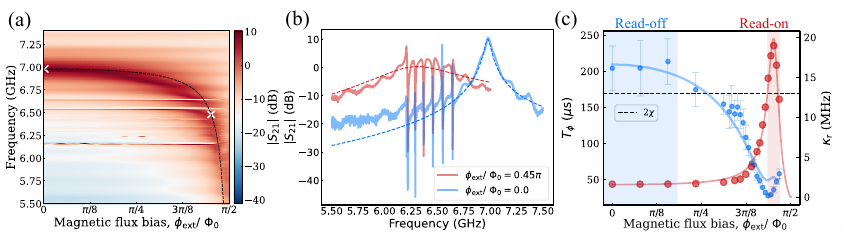}
    \caption{Characterization of tunable Purcell filter and photon-noise tolerance.    
    (a) Transmission spectrum $|S_{21}|$ of the Purcell filter as a function of SQUID magnetic flux bias. The dashed black curve indicates the fitted center frequency, with white “X” markers identifying the flux bias points corresponding to the configurations in Fig.~2(b). 
    (b) Transmission spectra with the filter tuned to resonance (red) and detuned (blue) relative to the readout resonators. Dashed lines represent simulated responses.
    (c) Measured resonator linewidth $\kappa_{\rm r}$ (red circles) and pure dephasing time $T_\mathrm{\phi}$ (blue circles) versus SQUID flux bias. Linewidths are extracted using $\chi$--$\kappa$ power spectroscopy~\cite{sank2025system}. Solid curves show theoretical fits, corresponding to a noise photon population of $\bar{n}_\mathrm{noise} = 5e^{-4}$. 
     \label{fig2}}
\end{figure*}

\textit{Tunable broadband Purcell filter}.---The experiments are conducted on a superconducting chip comprising six quarter-wavelength ($\lambda/4$) resonators coupled to a shared half-wavelength ($\lambda/2$) broadband Purcell filter, as illustrated in Fig.~\ref{fig1}(a).
The input and output ports are positioned at the shorted ends of the filter. 
A superconducting quantum interference device (SQUID) is integrated at the output end, with its inductance $L_{\rm S}(\phi_{\rm ext})$ controlled via a dedicated flux bias line~\cite{krantz2019quantum}, as depicted in the circuit model in Fig.~\ref{fig1}(b).
The resonance frequency $\omega_{\rm f}$ and quality factor $Q_{\rm f}$ of the Purcell filter are tunable through the SQUID inductance, following the relations~\cite{sandberg2008fast}:
\begin{equation}
\begin{aligned}
\omega_{\rm f} &= \frac{\omega_{\rm f0}}{1+L_{\rm S}/L_{\rm f0}}, \ 
Q_{\rm f} = \frac{R_0}{Z_{\rm f}\sin^2(\pi L_{\rm S}/L_{\rm f0})}.
\end{aligned}
\label{eq1}
\end{equation}
Here, $\omega_{\rm f0}$ and $L_{\rm f0}$ denote the bare frequency and effective inductance of the filter in the limit of zero SQUID inductance, while $R_0$ and $Z_{\rm f}$ represent the output port resistance and filter impedance, respectively. The filter linewidth is given by $\kappa_{\rm f} = \omega_{\rm f}/Q_{\rm f}$.
Figure~\ref{fig2}(a) presents the transmission response $|S_{21}|$ of the Purcell filter as a function of magnetic flux bias. The bare frequency $\omega_{\rm f0}$ is designed near $7~\mathrm{GHz}$, above the resonator frequencies, enabling tunability down into the resonator band for readout.
This tunability supports two operational regimes: a read-on regime with broader bandwidth ($\kappa_{\rm f}^{\rm on} \approx 900$~MHz) to accommodate a wider range of resonator frequencies, and a read-off regime with narrower band ($\kappa_{\rm f}^{\rm off} \approx 170$~MHz) to enhance qubit coherence protection~\cite{sete2015quantum}, as shown in Fig.~\ref{fig2}(b).
The effective resonator linewidth $\kappa_{\rm r}$ depends on the filter parameters as follows~\cite{sete2015quantum}:
\begin{equation}
\kappa_{\rm r} = \frac{4|\mathrm{g_{rf}}|^2}{\kappa_{\rm f}} \cdot \frac{1}{1 + \left[2(\omega_{\rm r} - \omega_{\rm f})/\kappa_{\rm f}\right]^2},
\end{equation}
where $\omega_{\rm r}$ is the resonator frequency and $g_{\mathrm{rf}}$ is the coupling strength between the resonator and the Purcell filter. 
A high-contrast $\kappa_{\rm r}$ is achieved, ranging from $2~\mathrm{MHz}$ to $20~\mathrm{MHz}$ (Fig.~\ref{fig2}(c)).
The measured ON/OFF ratio of 10 is consistent with the theoretical prediction $\kappa_{\rm r}^{\rm on}/\kappa_{\rm r}^{\rm off}=4(\omega_{\rm r} - \omega_{\rm f}^{\rm off})^2/(\kappa_{\rm f}^{\rm on}\kappa_{\rm f}^{\rm off})\approx9$, and is primarily limited by the minimum achievable SQUID inductance.

The measured pure dephasing time $T_2^{\phi}$ as a function of flux bias is presented in Fig.~\ref{fig2}(c), with the frequency-tunable transmon qubit biased at its flux sweet spot~\cite{koch2007charge}. Photon-noise-induced dephasing is strongly suppressed in the read-off regime, where $\kappa_{\mathrm{r}} \ll 2\chi$~\cite{Gambetta2006,yan2016flux}, resulting in a prolonged $T_{\phi}$ of 200~$\mu$s. In the read-on regime, the filter is tuned such that $\kappa_\mathrm{r}$ approaches $2\chi$, intentionally enhancing the qubit's sensitivity to resonator photons to enable faster readout~\cite{clerk2010introduction}. A fit to the dephasing rate yields a residual noise photon population of $5 \times 10^{-4}$. The seven-fold improvement in dephasing time between extreme values demonstrates the photon-noise tolerance of the tunable readout scheme. Furthermore, by optimizing $\kappa_\mathrm{r}$ to maximize the signal-to-noise ratio (SNR)~\cite{Gambetta2006,walter2017rapid}, this tunability enables rapid, high-fidelity readout.

\begin{figure}[t]
    \centering
    \includegraphics[width=\columnwidth]{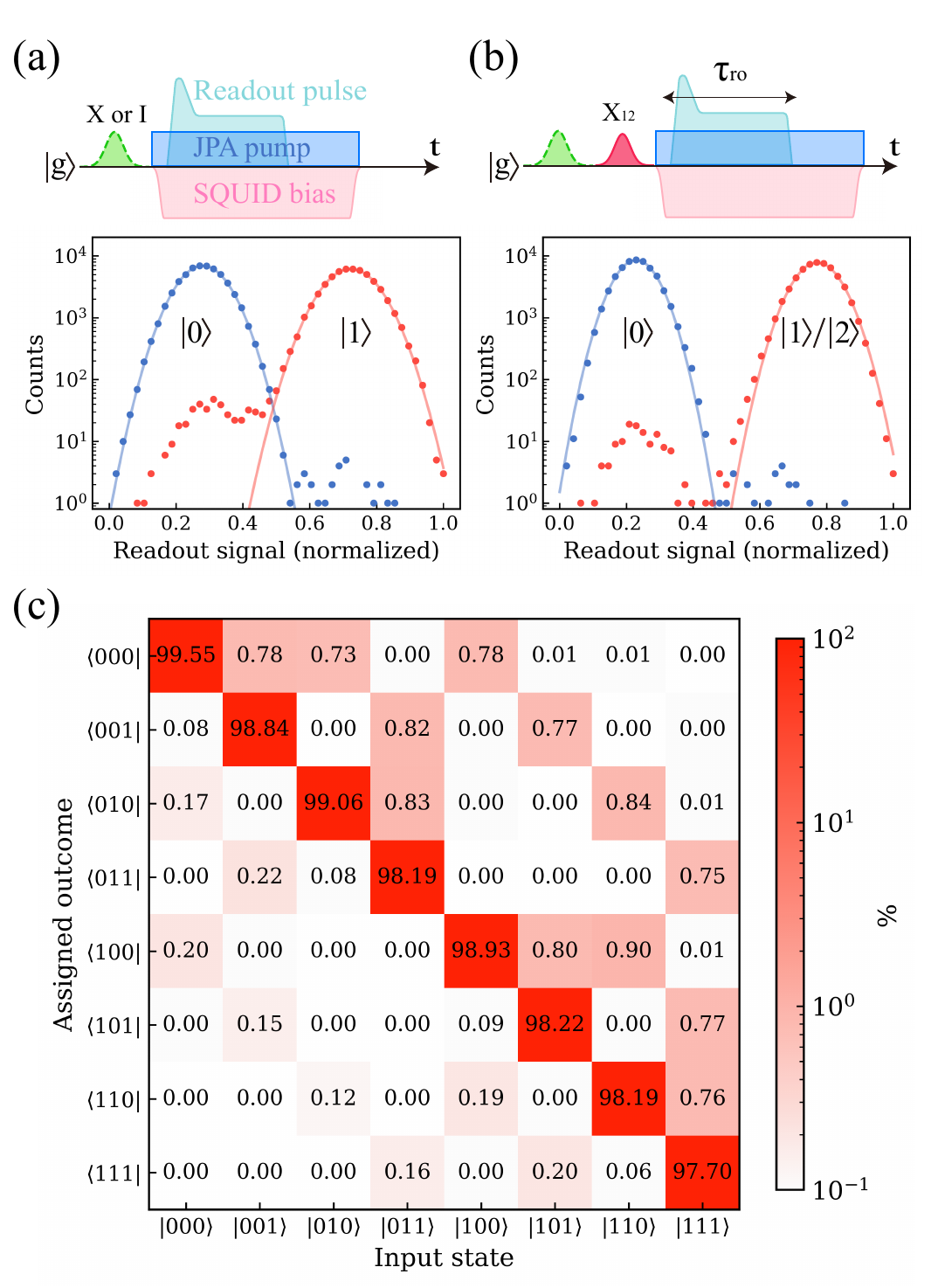}
    \caption{\raggedright Readout fidelity characterization.
    (a) Single-shot readout histograms of the integrated signal for Q2 using the pulse sequence shown above, with a 100~ns readout pulse. The JPA pump and filter SQUID bias signals are extended beyond the readout pulse to enable extended signal acquisition and resonator photon depletion. The total measurement length $\tau_{\rm m}$ is 250~ns, consisting of 25~ns pre-pulse and 125~ns post-pulse durations. The signal demodulation time $\tau_{\rm demod}$ is 200~ns. Solid lines represent Gaussian fits. The qubit ground state is prepared using a post-selection readout.
    (b) Readout histograms for Q2 utilizing the second excited state, with readout pulse length $\tau_{\mathrm{ro}} = 50$~ns and 120~ns demodulation time. The pulse sequence includes an additional $X_{12}$ gate (red) applied before readout to transfer $|1\rangle$ population to $|2\rangle$. 
    (c) Assignment matrix for simultaneous three-qubit readout using 100~ns readout pulses, averaged over 60,000 shots. The states are labeled as $|{\rm Q}1,{\rm Q}2,{\rm Q}3\rangle$.
     \label{fig3}}
\end{figure}

\begin{figure}[t]
    \centering
    \includegraphics[width=\columnwidth]{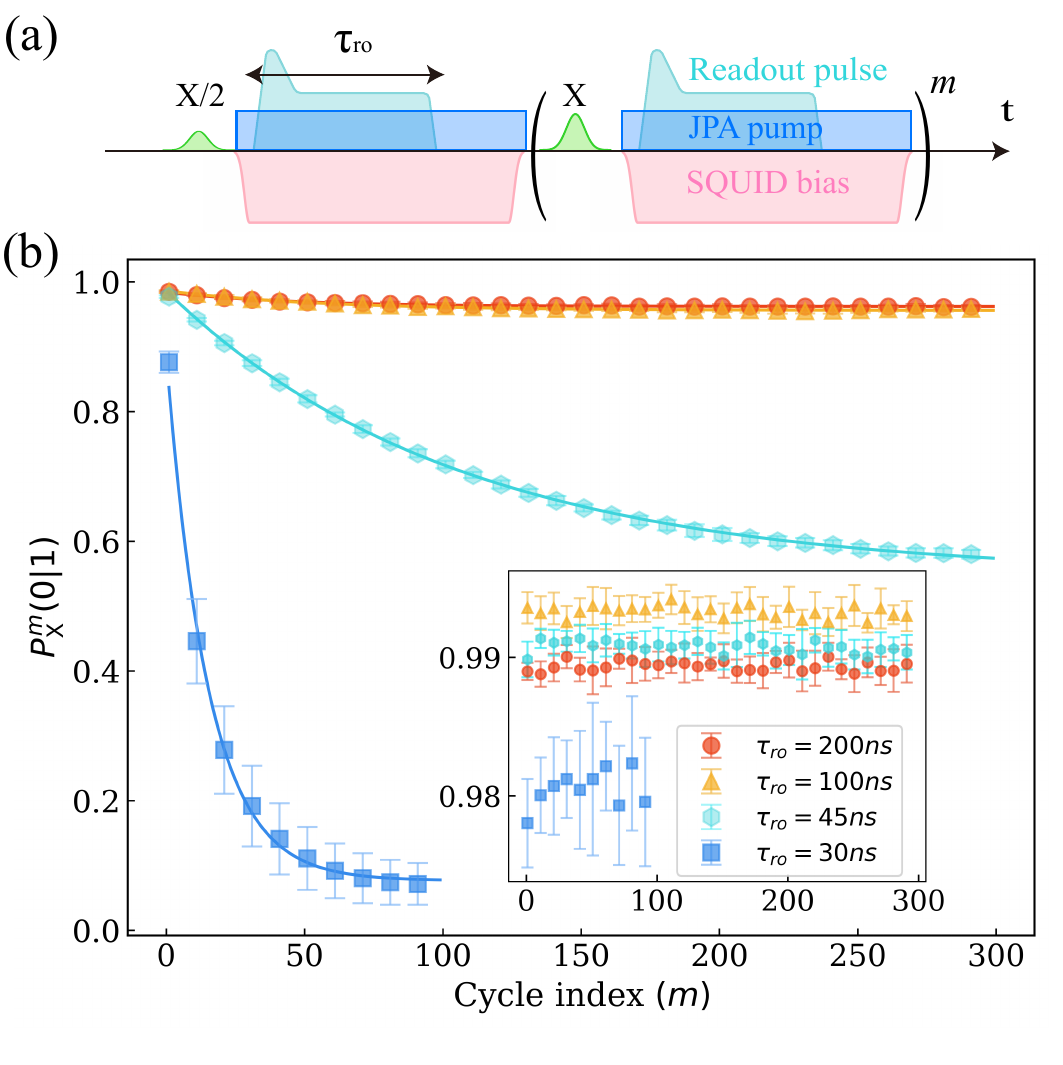}
    \caption{\raggedright Benchmarking readout-induced leakage using repeated measurements.
    (a) $\pi$-QND sequence for quantifying readout-induced leakage. An initial $X/2$ gate prepares an equal superposition of $\ket{0}$ and $\ket{1}$, followed by interleaved $X$ gates between measurement pulses to alternately flip the qubit state.
    (b) Measured leakage probabilities $P^{m}_{X}(0|1)$ (dots) versus cycle index $m$ for different readout pulses. Solid curves represent exponential fits. The inset displays the corresponding probabilities for $P^{m}_{X}(1|0)$.
     \label{fig4}}
\end{figure}

\textit{High performance readout}.---We demonstrate rapid, high-fidelity, and low back-action readout in the read-on regime.  
The pulse sequence is shown in Fig.~\ref{fig3}(a), where a filter SQUID bias pulse and a pump signal for Josephson parametric amplifier (JPA) are applied concurrently with the readout pulse. The qubit is initialized in the ground state via a post-selection readout before the sequence. 
The filter bias amplitude is optimized to maximize the SNR, and its duration exceeds that of the readout pulse to accelerate resonator photon depletion~\cite{bultink2016active}. Owing to the large $\kappa_{\rm r}^{\rm on}~\approx 16$~MHz during readout, a 100~ns depletion time ($\sim 10 /\kappa_{\rm r}^{\rm on}$) suffices for resonator reset. We find that adding a 25~ns rising edge to the bias pulse improves fidelity, likely due to reduced flux pulse distortion. Consequently, the total measurement time $\tau_{\rm m}$ extends 150~ns beyond the readout pulse length $\tau_{\rm ro}$.
Using a 100~ns readout pulse, the readout fidelity for Q2, defined as $\mathcal{F_{\rm R}} = {[P(0|g) + P(1|e)]}/{2}$ is measured to be 99.61\%. Here, $P(0|g)$ ($P(1|e)$) denotes the probability of measuring state $|0\rangle$ ($|1\rangle$) when the qubit is prepared in $|g\rangle$ ($|e\rangle$). This fidelity is primarily limited by relaxation error during readout. The effective $T_1^{\rm r}$ under readout is measured to be 26~$\mu$s~\cite{sm2025,thorbeck2024readout}, and the relaxation error is estimated as $\tau_{\rm demod} / (2 T_1^{\rm r}) = 0.38\%$, where $\tau_{\rm demod} = 200$~ns is the demodulation time.
To mitigate relaxation error, we apply an ${\rm X}_{12}$ gate before readout to pre-excite the $|1\rangle$ population to $|2\rangle$, as shown in Fig.~\ref{fig3}(b). Employing the multilevel readout technique~\cite{elder2020high,wang2021optimal}, we achieve a readout fidelity of 99.87\% using a 50~ns readout pulse and a demodulation time of 120~ns.

The tunable broadband Purcell filter is compatible with multiplexed readout, as its saturation power significantly exceeds typical readout power~\cite{sm2025}.
Simultaneous 100~ns readout pulses applied to three qubits yield an average fidelity of 99.53\%, comparable to the average individual fidelity of 99.58\%. 
Readout crosstalk is characterized by preparing the eight basis states of three qubits and measuring the assignment probabilities. The average cross-fidelity is 0.02\%~\cite{heinsoo2018rapid}, extracted from the probability matrix in Fig.~\ref{fig3}(c). The matrix shows that infidelities primarily arise from relaxation errors (values above the diagonal), while values below the diagonal indicate measurement-induced excitations.

Measurement-induced leakage is characterized via repeated measurements using the sequence shown in Fig.~\ref{fig4}(a). 
An initial $X/2$ gate prepares an equal superposition of $|0\rangle$ and $|1\rangle$, and an $X$ gate is inserted between readout pulses to alternately flip the qubit state. 
We define $P^{m}_{X}(0|1)$ ($P^{m}_{X}(1|0)$) as the probability of obtaining $\ket{0}$ ($\ket{1}$) in the $m$-th measurement given that the $(m-1)$-th result was $\ket{1}$ ($\ket{0}$).
The probabilities are plotted in Fig.~\ref{fig4}(b) for different readout pulse lengths.
$P^m_X(0|1)$ exhibits exponential decay with $m$, while $P^m_X(1|0)$ shows no decay (inset), since leakage states in transmon qubit produce signals closer to $|1\rangle$~\cite{wang2021optimal,sank2016measurement,khezri2023measurement}.
The decay of $P^{m}_{X}(1|0)$ can be modeled by~\cite{sm2025}:
\begin{equation}
P^{m}_{X}(0|1)=A(1-\mathcal{L}_\downarrow - \mathcal{L}_\uparrow)^m+B.
\label{leakage_eq1}
\end{equation}
where $\mathcal{L}_\uparrow$ is the leakage rate from the computational subspace and $\mathcal{L}_\downarrow$ is the seepage rate from leakage states back into the computational subspace. 
The steady-state value $B$ reflects the balance between leakage and seepage processes: $B \mathcal{L}_\uparrow = (1 - B) \mathcal{L}_\downarrow$.
We extract $\mathcal{L}_\downarrow$ and $\mathcal{L}_\uparrow$ by fitting the measured probabilities using Eq.~\ref{leakage_eq1}.
The leakage rate is averaged over $\ket{0}$ and $\ket{1}$, as the qubit state is alternatively flipped during the experiment. 
As summarized in Table~\ref{table 1}, $\mathcal{L}_\uparrow$ increases with the readout power (and resonator photon number), whereas the seepage rate decreases with readout power. This trend is consistent with theoretical predictions that the qubit state is more likely to leak into higher energy levels with increased resonator photon populations~\cite{dumas2024measurement,shillito2022dynamics}, leading to longer decay times for leakage states to return to the computational subspace.
The leakage benchmarking method yields leakage rates consistent with those obtained via the randomized benchmarking approach introduced in Ref.~\cite{hazra2025benchmarking}, as detailed in the supplementary material~\cite{sm2025}.
With a 100~ns pulse, we achieve a low leakage rate of 0.08\% and a high QND fidelity of 99.37\%~\cite{spring2025fast,sm2025}, demonstrating that low back-action readout is compatible with the tunable filter.

\begin{table}[t]
\centering
\begin{tabular}{p{3.8cm} p{1.0cm} p{1.0cm} p{1.0cm} p{1.0cm} }
\hline 
\hline 
Readout Duration (ns) & 200 & 100 &  45  & 30 \\ 
\hline 
Photon number, $\bar n_{r0}$& 3.4 & 9.6  &  13.8 & 21.5 \\ 
Photon number, $\bar n_{r1}$& 0.8 & 1.8  &  2.7 & 5.0 \\ 
Readout fidelity, $\mathcal{F_{\rm R}}$ (\%) &  99.56 & 99.58   &  99.52 & 99.23  \\ 
QND fidelity, $\mathcal{F_{\rm Q}}$ (\%) &  99.15  & 99.37 &  99.37  & 98.79  \\ 
Seepage rate, $\mathcal{L}_\downarrow$ (\%) &  2.12  & 1.70  &  0.51  & 0.49  \\ 
Leakage rate, $\mathcal{L}_\uparrow$ (\%) &  0.08  & 0.08 &  0.43  & 5.91  \\ 
\hline
\end{tabular}
\caption{
Readout performance of Q2 for different pulse durations. $\bar{n}_{r0}$ and $\bar{n}_{r1}$ denote the average resonator photon number for qubit states $\ket{0}$ and $\ket{1}$, respectively. The QND fidelity is defined as $\mathcal{F}_{\rm Q} = \left[P(0|0) + P(1|1)\right] / 2$, where $P(i|i)$ represents the probability of measuring state $\ket{i}$ following a previous measurement of $\ket{i}$.
}
\label{table 1}
\end{table}

\begin{figure}[t]
    \centering
    \includegraphics[width=\columnwidth]{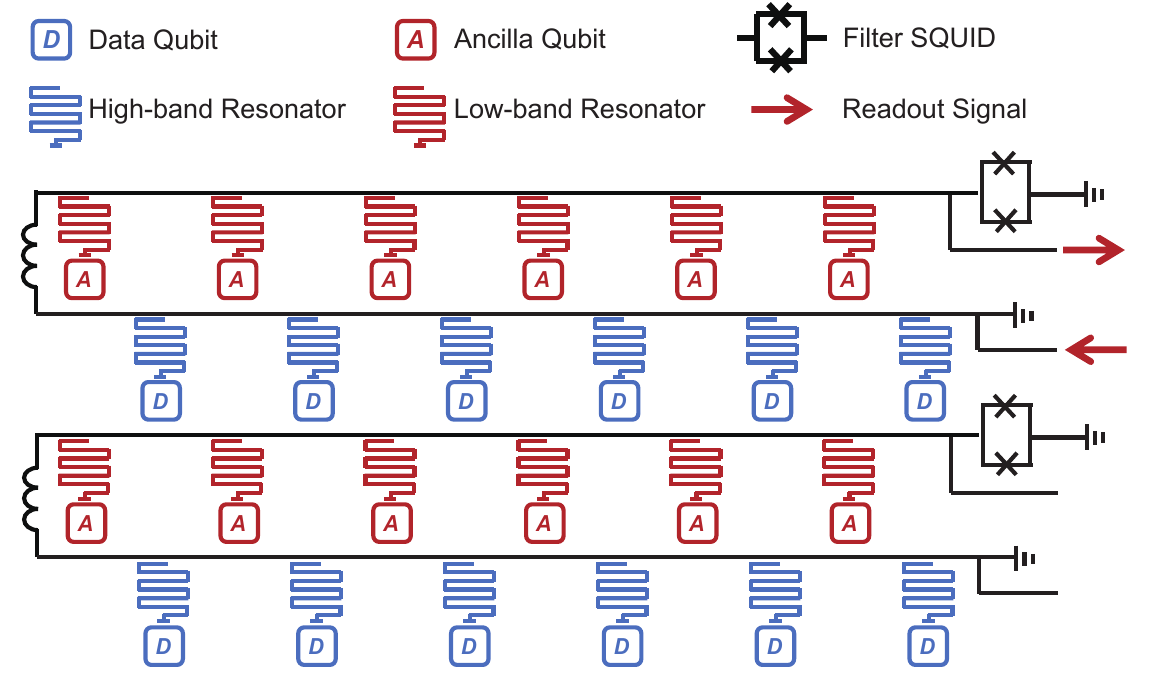}
    \caption{\raggedright Multiplexed readout architecture based on a tunable broadband Purcell filter. Data qubits (blue squares) and ancilla qubits (red squares) share a common $\lambda/2$ Purcell filter, with their resonators operating in spectrally separated frequency bands. Selective measurement of ancilla (data) qubits is achieved by tuning the filter frequency to the corresponding band, while data (ancilla) qubits are protected from measurement-induced dephasing due to suppressed resonator linewidths.
     \label{fig5}}
\end{figure}

\textit{Scalable multiplexed readout for surface code}.---We now discuss the scalability of the tunable broadband filter for QEC. Enhancing readout multiplexing capacity is crucial for scaling superconducting quantum processors, as it reduces the number of required cryogenic amplifiers and circulators.
A proposed 2D surface code layout incorporating the tunable broadband Purcell filter is illustrated in Fig.~\ref{fig5}. In the design, resonators for both data and ancilla qubits are coupled to a common tunable filter while occupying distinct frequency bands.
The tunable filter enables selective measurement of data or ancilla qubits by dynamically shifting its frequency into the target band---a strategy made possible by the fact that surface code protocols typically do not require simultaneous measurement of both qubit types~\cite{fowler2012surface} .
In the read-off regime, the filter is detuned from both frequency bands to provide robust protection against photon noise.
During readout of ancilla (or data) qubits, the filter frequency is aligned to the corresponding band, optimizing the resonator linewidth to match the dispersive shift and thereby maximize the SNR. For unmeasured qubits, the filter remains detuned from their resonator band, resulting in reduced resonator linewidths~\cite{sm2025}.
The resulting $\kappa$-$2\chi$ mismatch and interband detuning effectively protect unmeasured qubits from both thermal noise and parasitic measurement-induced dephasing~\cite{Gambetta2006,blais2021circuit,heinsoo2018rapid}.
Notably, the architecture remains compatible with narrower-bandwidth JPAs~\cite{mutus2014strong,white2023readout,frattini2018optimizing}, as data qubits can be measured outside the amplification bandwidth by exploiting multilevel readout techniques~\cite{wang2025longitudinal,chen2023transmon}.  
Compared to conventional surface code readout designs~\cite{google2025quantum,gao2025establishing}, this approach effectively doubles the multiplexing capacity.

\textit{Conclusions}.---We overcome the inherent trade-off between readout speed and photon-noise-induced dephasing by dynamically controlling resonator linewidths using a tunable broadband Purcell filter. In the read-off regime, the resonator linewidth $\kappa_{\mathrm{r}}$ is suppressed to values much smaller than the dispersive shift $\chi$, enhancing the qubit dephasing time by a factor of 7 compared to the read-on regime.
In the read-on regime ($\kappa_{\rm r} \approx 2\chi$), we achieve high performance readout, including: (1) 99.6\% 
fidelity with 100~ns readout pulses, limited primarily by qubit relaxation time($T_1$); (2) 99.9\% fidelity using 50~ns pulses via a multilevel readout protocol; and (3) high QND performance with 99.4\% fidelity and leakage rates below 0.1\% for 100~ns readout.
The multiplexing capability of our nonlinear Purcell filter is validated through simultaneous readout of three qubits, achieving an average fidelity of 99.5\% with minimal crosstalk, consistent with $T_1$-limited performance.
Building on these results, we further propose a multiplexed readout architecture for surface-code-based QEC with enhanced multiplexing capacity, showing the potential of tunable Purcell filters to enable scalable QEC implementation.

\textit{Acknowledgement}.---We thank Wei Dai for insightful discussions on the readout-induced leakage benchmarking method.
This work was supported by the Innovation Program for Quantum Science and Technology (2021ZD0301703), the Science, Technology and Innovation Commission of Shenzhen Municipality (KQTD20210811090049034), the National Natural Science Foundation of China (123b2071, 12174178, 12374474).

\bibliography{main}
\end{document}


\title{Supplementary Information for ``High-performance multiplexed readout of superconducting qubits with a tunable broadband Purcell filter''}


\maketitle

\setcounter{equation}{0}
\setcounter{figure}{0}
\setcounter{table}{0}
\setcounter{page}{1}

\renewcommand{\theequation}{S\arabic{equation}}
\renewcommand{\thefigure}{S\arabic{figure}}
\renewcommand{\thetable}{S\arabic{table}}

\newcommand{\figtitle}[1]{\textbf{#1}}
\newcommand{\subfiglabel}[1]{\textbf{#1}}

\tableofcontents

\clearpage

\section{Theoretical model of a tunable Purcell filter}

The Purcell filter used in our experiments is a shorted half-wavelength ($\lambda/2$) coplanar waveguide resonator, made tunable by incorporating a SQUID at one of its shorted ends.
By adjusting the flux bias ($\phi_{\rm ext}$) applied to the SQUID, the effective inductance $L_{\mathrm{S}}$ of the SQUID is modified, thereby tuning the filter frequency $\omega_{\rm f}$ according to~\cite{sandberg2008fast}
\begin{equation}
\omega_{\rm f}(\phi_{\rm ext})=\omega_{\rm f0}/[1+L_{\mathrm{S}}(\phi_{\rm ext})/L_{\rm f0}],
\end{equation}
where
\begin{equation}
L_{\mathrm{S}}(\phi_{\rm ext})=\frac{\Phi_0}{4\pi\\I_{\rm c}\sqrt{\cos^2(\pi\phi_{\rm ext}/\Phi_0)}}.
\end{equation}
Here $\omega_{\rm f0}$ and $L_{\rm f0}$ denote the filter frequency and total inductance in the limit of negligible SQUID inductance, respectively, and $I_{\rm c}$ is the critical current of the SQUID.

The total external loss of the filter comprises the losses through both the input and output ports. The overall quality factor $Q_{\rm f}$ is given by:
\begin{equation}
\frac1{Q_{\rm f}(\phi_{\rm ext})}=\frac1{Q_{\rm fout}(\phi_{\rm ext})}+\frac1{Q_{\rm fin}},
\end{equation}
where $Q_{\rm fout}$ and $Q_{\rm fin}$ represent the quality factors associated with the output and input ports, respectively.
These quality factors depend on the electrical positions of the ports, which follow the sinusoidal voltage distribution along the half-wavelength resonator.
Since the input port coupling is significantly weaker than the output port coupling, the total quality factor is dominated by the output port: $Q_{\rm f} \approx Q_{\rm fout}$.
Placing the SQUID between the readout port and the shorted end causes the electrical position of the output port to vary with the SQUID inductance, modulating $Q_{\rm fout}$ as~\cite{pozar2021microwave}
\begin{equation}
{Q_{\rm fout}(\phi_{\rm ext})}=\frac{Q_{\rm f0}}{\sin^2\left(\pi L_{\mathrm{S}}(\phi_{\rm ext})/L_{\rm u}l_{\rm p}\right)}.
\end{equation}
Here, $l_{\rm p}$ is the physical length of the filter, $L_{\rm u}$ is the per-unit-length inductance of the filter, and $Q_{\rm f0}=R_{\rm 0}/Z_{\rm f}$ is a constant determined by the characteristic impedance of the filter $Z_{\rm f}$ and the port resistance $R_{\rm 0}$~\cite{pozar2021microwave}.
As $L_{\rm S}(\phi_{\rm ext})$ increases, the electrical position of the output tap shifts closer to the current antinode at the $\pi/2$ point, where the voltage amplitude is maximized. This enhances the coupling strength to the output line, thereby reducing the external quality factor $Q_{\rm f}$.

\section{Experimental setup and device parameters}

\begin{figure}[b]
    \centering
    \includegraphics[width=0.9\linewidth]{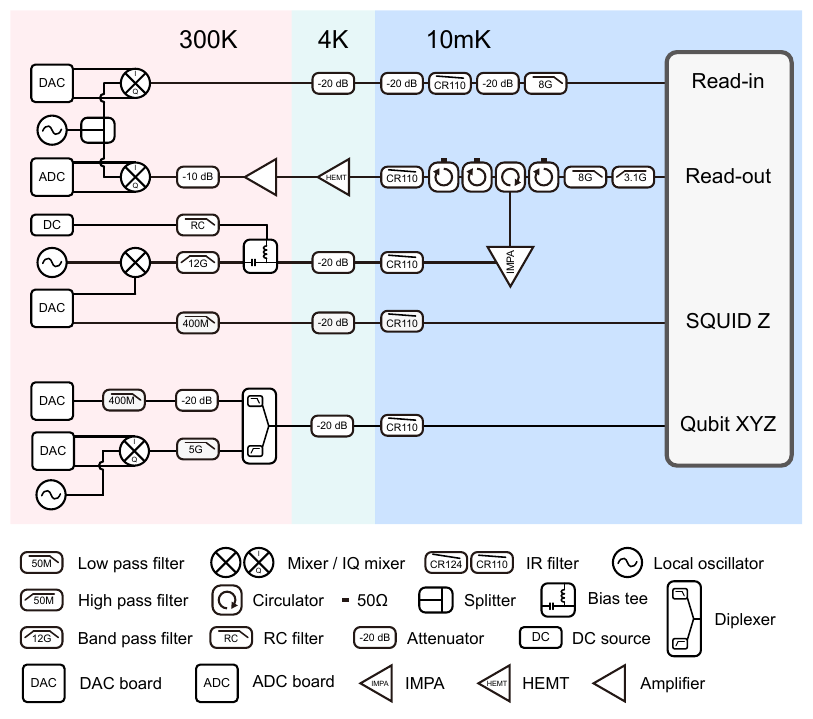}
    \caption{Schematic of the experimental setup. }
    \label{setup}
\end{figure}

\begin{figure}[t]
    \centering
    \includegraphics[width=0.4\linewidth]{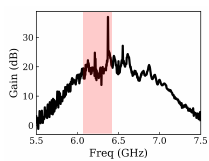}
    \caption{{Gain profile of the Josephson parametric amplifier (JPA) used in the experiments. The colored region marks the readout frequency range for the three qubits used in the experiments.} }
    \label{JPA}
\end{figure}

\begin{table*}[t]
\centering
\begin{tabular}{p{1cm} p{1.8cm} p{1.8cm} p{1.8cm} p{1.8cm} p{1.5cm} p{1.5cm} p{1.5cm} p{1.5cm}}
\hline 
\hline 
&$\omega_{\rm q}~\mathrm{(GHz)}$ & $\omega_{\rm r}~\mathrm{(GHz)}$ & 2$\chi~\mathrm{(MHz)}$ & $\alpha~\mathrm{(MHz)}$ & $T_{1}~\mathrm{(\mu{s})}$ &  $T_{2}^{\mathrm{*}}~\mathrm{(\mu{s})}$ &$T_{2}^{\mathrm{echo}}~\mathrm{(\mu{s})}$&~~$\eta$ \\ 
\hline 
Q1&4.696 & 6.284 & 15 & -196 & 47.24  &  30.63 & 49.43 & 0.18 \\ 
Q2&4.615 & 6.362 & 13 & -206 & 47.78  &  9.39 & 68.31 & 0.29 \\
Q3&4.681 & 6.449 & 13 & -193 & 36.11  &  17.71 & 45.74 & 0.32 \\ 
\hline
\end{tabular}
\caption{Parameters of the qubits used in the experiment. $\eta$ denotes the readout efficiency~\cite{bultink2018general}. }
\label{qubits parameters}
\end{table*}

The quantum processor is mounted on the mixing chamber plate of a dilution refrigerator and operates at a base temperature of 10~mK. A schematic of the full experimental setup is shown in Fig.~\ref{setup}. Control signals are generated and measurement data acquired using a Microwave Measurement and Control System (M2CS)~\cite{zhang2024m2cs}. The gain profile of the Josephson parametric amplifier (JPA)~\cite{mutus2014strong,grebel2021flux} is presented in Fig.~\ref{JPA}, with the colored region highlighting the readout frequency band for the three qubits. Table~\ref{qubits parameters} summarizes the characteristic parameters of the three qubits used in the experiments.

\begin{figure}[b!]
    \centering
    \includegraphics[width=0.9\linewidth]{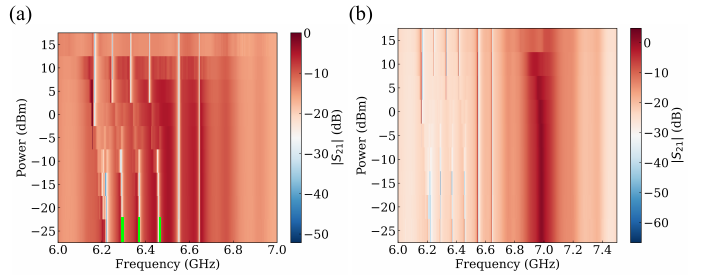}
    \caption{ Measured transmission of readout line versus VNA power in read-on (a) and read-off (b) regimes. The green lines in (a) indicate the readout resonators used in the experiment.
    The filter bandwidth increases significantly at high  power ($>$5~dBm), which is well above the typical readout power used (-20~dBm).}
    \label{filter_power}
\end{figure}

\section{High-power response of tunable Purcell filter}

At high readout powers, the current across the filter SQUID may exceed its critical current, causing the Josephson junctions to become resistive and degrade readout performance. Figure~\ref{filter_power} shows the measured response of the readout line as a function of the output power from the vector network analyzer (VNA).
For powers exceeding approximately 5~dBm, we observe a significant reduction in the filter quality factor, suggesting that the current may approach or exceed the SQUID's critical current. We note, however, that this power level is substantially higher than that used during typical readout, which is around -20~dBm.
We further demonstrate theoretically that the current in the SQUID remains well below the critical current during multiplexed readout. The relationship between the average photon number in the filter, $\bar n_{\rm f}$, and that in the readout resonator, $\bar n_{\rm r}$, is given by~\cite{sunada2024photon}:
\begin{equation}
\bar n_{\rm f}=(\frac{\Delta_{\rm rd}}{ g_{\rm rf}})^2 \bar n_{\rm r},
\end{equation}
where $\Delta_{\rm rd}$ is the detuning between resonator frequency and the probe frequency, and $g_{\rm rf} \approx 50$~MHz is coupling strength between the readout resonator and the Purcell filter.
Choosing the probe frequency at the midpoint between the resonator frequencies corresponding to the qubit's ground and excited states, $\omega_{\rm d}=(\omega_{\rm r}^{\ket{0}}+\omega_{\rm r}^{\ket{1}})/2$, yields $\Delta_{\rm rd}^2=\chi^2$, where $\chi$ is the dispersive shift.
Assuming a conservatively large photon number of $\bar n_{\rm r}=50$ in the readout resonator---well beyond typical values of approximately 5 photons used in low-leakage quantum nondemolition (QND) measurements---the corresponding average photon number in the filter is $\bar n_{\rm f}=0.72$. The relationship between the average photon number and the filter current $I_{\rm f}$ is derived from the energy stored in the filter circuit~\cite{krantz2019quantum,sunada2024photon} 
\begin{equation}
\bar n_{\rm f}\hbar\omega_{\rm f}= \frac{I^2}{4 \omega_{\rm f}^2 C_{\rm f}},
\end{equation}
where $C_{\rm f}$ is the effective capacitance of the filter. For $\bar{n}_{\rm f} = 0.72$ and $\omega_{\rm f} / (2\pi) = 6$~GHz, the calculated current is $I_{\mathrm{f}} = 18.5$~nA.
Given a typical SQUID inductance of $L_{\rm S} = 0.5$~nH during readout, the critical current is estimated as $I_{\mathrm{c}} = \Phi_0 / (2\pi L_{\rm S}) \approx 0.66~\mu$A, where $\Phi_0$ denotes the flux quantum. The operational current of $18.5~\mathrm{nA}$ is more than an order of magnitude lower than this critical threshold. 
This substantial safety margin confirms that the tunable filter design robustly supports simultaneous readout of multiple qubits without compromising the superconducting state of the SQUID.


\section{Purcell $T_1$ protection}

\begin{figure}[b!]
    \centering
    \includegraphics[width=0.9\linewidth]{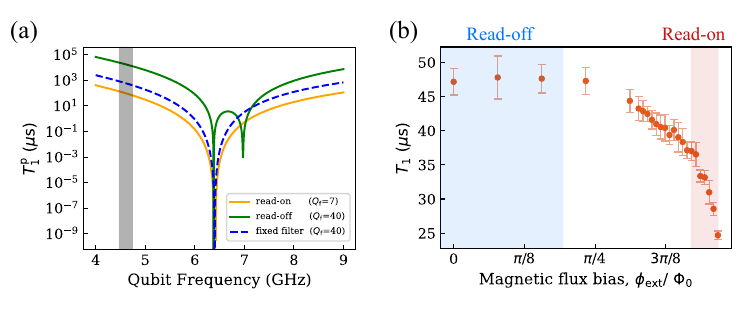}
    \caption{Purcell $T_1$ protection provided by the tunable broadband Purcell filter.
    (a) Estimated Purcell limited $T_1$ using Eq.~\ref{purcell_T1}. The blue dashed line represents a fixed-frequency Purcell filter with $Q_{\rm f}=40$; the green line corresponds to the tunable filter in the read-off regime; the orange line shows the tunable filter in the read-on regime with $Q_{\rm f}=7$. 
    In the read-off regime, the tunable filter frequency is detuned from the resonator band, resulting in reduced $\kappa_{\rm r}$ and enhanced qubit $T_1$ protection.
    The dark shaded region indicates the qubit frequency range used in this experiment.
    (b) Experimentally measured relaxation time of Q2 as a function of the SQUID flux bias. Reduced $T_1$ protection is observed in the read-on regime due to the lower filter quality factor $Q_{\rm f}$.
    }
    \label{T1_protection}
\end{figure}

Similar to conventional linear Purcell filters, the tunable design also provides protection for the qubit relaxation time. The Purcell-limited relaxation time $T_{1}^{\rm p}$ is given by~\cite{jeffrey2014fast}
\begin{equation}
\kappa_\mathrm{eff}T_{1}^{\rm p}=4\frac{\Delta_{\rm qr}^2\Delta_{\rm qf}^2}{\mathrm{g_{qr}}^2\omega_{\rm q}^2/Q_{\rm f}^2},
\label{purcell_T1}
\end{equation}
where $\Delta_{\rm qr} = \omega_{\rm q}-\omega_{\rm r}$ and $\Delta_{\rm qf} = \omega_{\rm q}-\omega_{\rm f}$ denote the qubit-resonator and qubit-filter detuning, respectively, and $\mathrm{g_{qr}}$ represents the coupling strength between the qubit and the readout resonator. 
The Purcell-limited $T_1$ depends critically on both the filter quality factor $Q_{\rm f}$ and the resonator linewidth $\kappa_{\rm r}$.
In the read-off regime, the combination of reduced $\kappa_{\rm r}$ and increased $Q_{\rm f}$ significantly enhances the Purcell-limited $T_1$, as shown in Fig.~\ref{T1_protection}(a).
Compare to fixed-frequency Purcell filters with the same quality factor, the reduced $\kappa_{\rm r}$ resulting from detuning between the resonator band and the tunable filter frequency provides superior $T_1$ protection.
However, in the read-on regime, the reduced quality factor ($Q_{\rm f} = 7$) diminishes coherence protection. This observation is confirmed by experimental measurements of Q2's $T_1$ as a function of flux bias, presented in Fig.~\ref{T1_protection}(b). In the read-off regime, the qubit $T_1$ remains stable and is primarily limited by intrinsic losses. In contrast, the read-on regime shows a Purcell-limited $T_1$ of approximately 25$\mu$s, attributable to the insufficient filter quality factor. 
We note that increasing $Q_{\rm f}$ to 14 in the read-on regime could elevate the Purcell-limited $T_1$ beyond 100~$\mu$s. For enhanced scalability, incorporating a band-stop filter~\cite{zhou2024high} or an intrinsic Purcell filter~\cite{sunada2022fast} could provide higher-order $T_1$ protection.

\section{Readout fidelity benchmarking}

\begin{figure}[b!]
    \centering
    \includegraphics[width=0.9\linewidth]{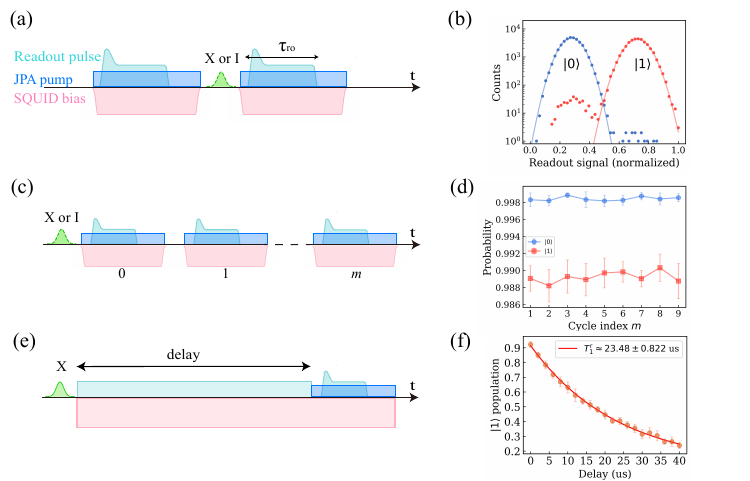}
    \caption{
    (a) Pulse sequence for benchmarking readout fidelity. The first readout pulse is used to post-select the qubit into the ground state.
    (b) Histograms of the integrated readout signals for the 45~ns single-shot readout of Q2, obtained using the pulse sequences in (a).
    (c) Pulse sequence for benchmarking QND fidelity. Measurement pulses are applied repetitively, with a 40~ns idle time inserted between them. The qubit is prepared in $|g\rangle$ ($|e\rangle$) using an I (X) gate prior to the sequence.
    (d) Probability of measuring state $\ket{i}$ for Q2 following a previous measurement of $\ket{i}$ versus cycle index, using the pulse sequence in (c). The readout pulse length is 45~ns. Error bars denote the standard deviation from 10 repeated experiments, each consisting of 4000 readout shots.
    (e) Pulse sequence for benchmarking effective $T_{1}$ during readout. The qubit is prepared in the excited state with an X gate, followed by a varied duration with filter SQUID bias and readout pulse applied. 
    (f) Experimental measured effective $T_{1}^{\rm r}$ of Q2 during readout, using the pulse sequence in (e). 
    }
    \label{single_readout}
\end{figure}

We compare the experimental sequence for measuring readout fidelity and QND fidelity in Fig.~\ref{single_readout}. 
Fig.~\ref{single_readout}(a) shows the experimental sequence for measuring readout fidelity. The first readout pulse is used to post-select the qubit into the ground state. The readout fidelity is defined as $\mathcal{F_{\rm R}} = {(P(0|g) + P(1|e))}/{2}$, where $P(0|g)$ ($P(1|e)$) is the probability of measuring state $|0\rangle$ ($|1\rangle$) when the qubit is prepared in $|g\rangle$ ($|e\rangle$). Histograms of the integrated readout signals for 45~ns single-shot readout of Q2 are shown in Fig.~\ref{single_readout}(b), yielding a readout fidelity of 99.52\%.

We assess QND readout fidelity through repeated measurements using the pulse sequence shown in Fig.~\ref{single_readout}(c). To ensure resonator photon depletion, the JPA pump signal and the filter SQUID bias are maintained for a duration longer than the readout pulse itself, followed by an additional 40 ns idle period before the next measurement.
The QND fidelity for the $m$-th readout cycle is defined as $\mathcal{F}^m_{\rm QND} = {[P^{m}(0|0) + P^{m}(1|1)]}/{2}$, where $P^{m}(i|i)$ is the conditional probability of obtaining result $\ket{i}$ in the $m$-th measurement, given that the $(m-1)$-th measurement result was $\ket{i}$.
The probabilities $P^{m}(0|0)$ and $P^{m}(1|1)$ are plotted in Fig.~\ref{single_readout}(d), demonstrating that the QND fidelity $\mathcal{F}^m_{\rm QND}$ remains stable over all ten measurement cycles.
Individual readout and QND fidelities for the three qubits, obtained using 45~ns readout pulses, are summarized in Table~\ref{readout performance}. Here, $\bar{n}_{r0}$ ($\bar{n}_{r1}$) denotes the average resonator photon population when the qubit is prepared in state $\ket{0}$ ($\ket{1}$). 
We compare the QND fidelity of simultaneous readout of Q2 and Q3 with the individually measured values in Table~\ref{multi_QND_fid}. The QND fidelity under simultaneous readout shows no significant degradation compared to individual readout.

\begin{table*}[h]
\centering
\begin{tabular}{p{1.5cm} p{3cm} p{3cm} p{3cm} p{3cm}}
\hline 
\hline 
 & Duration & Readout fidelity & QND fidelity & $\bar{n}_{r0}$/$\bar{n}_{r1}$ \\ 
\hline 
Q1 &45~ns & 99.56\%  & 99.24\% & 17.3/2.0  \\ 
Q2 &45~ns & 99.52\% & 99.37\% & 13.8/2.7  \\
Q3 &45~ns & 99.34\% & 99.14\% & 8.5/3.8  \\ 
\hline
\end{tabular}
\caption{
Single-shot readout and QND performance. Fidelities are measured individually using a 45~ns readout pulse. The demodulation time $\tau_{\mathrm{demod}}$ and total measurement time $\tau_{m}$ are fixed at 200~ns for all measurements. The demodulation window is optimized to maximize the SNR~\cite{bengtsson2024model}.
}
\label{readout performance}
\end{table*}

\begin{table*}[t]
\centering
\begin{tabular}{p{3cm} p{2cm} p{2cm} p{2cm} p{2cm} p{2cm} p{2cm}}
\hline 
\hline 
 & \textbf{Q2} & \textbf{Q3}  & $\textbf{Q2}^{*}$  & $\textbf{Q3}^{*}$    \\ 
\hline 
QND fidelity& $99.33\%$ & $99.44\%$  &  $99.31\%$ & $99.45\%$ \\
\hline
\end{tabular}
\caption{QND fidelity unders simultaneous readout of Q2 and Q3. Columns marked with (without) an asterisk (*) represent results from simultaneous (individual) readout. The readout pulse length $\tau_{\mathrm{ro}}$, demodulation time $\tau_{\mathrm{demod}}$, and total measurement time $\tau_{m}$ are 100~ns, 200~ns, and 250~ns, respectively.}
\label{multi_QND_fid}
\end{table*}

\section{Readout error budget}

To analyze the readout error, we characterize the effective qubit relaxation time $T_1^{\mathrm{r}}$ during readout using the pulse sequence shown in Fig.~\ref{single_readout}(e). After initializing the qubit with an X gate, we apply a probe pulse with amplitude matching the readout pulse and variable duration, together with the filter SQUID bias. 
The measured relaxation time for Q2 during readout is presented in Fig.~\ref{single_readout}(f). Our results indicate a readout-induced suppression of the qubit lifetime~\cite{thorbeck2024readout}.
The contribution of relaxation error to the total readout error is given by $0.5(1-e^{-\tau/T_{1}^{\rm r}}) \approx \tau/(2T_{1}^{\rm r})$, where $\tau$ denotes the demodulation time.
Table~\ref{erroranalysis} summarizes the error budget for the three qubits using 100~ns readout pulses. The separation error is calculated as $\epsilon_{\mathrm{sep}} = \frac{1}{2}\mathrm{erfc}(\sqrt{\mathrm{SNR}}/2)$~\cite{krantz2019quantum}.
Due to the low saturation power of the JPA~\cite{mutus2014strong,macklin2015near,frattini2018optimizing,white2023readout}, rapid simultaneous readout of all three qubits leads to a degraded SNR, thereby increasing the separation errors. The total readout error is dominated by relaxation error. We note that the estimated relaxation error is subject to considerable uncertainty, since the intra-cavity photon population varies dynamically during the readout process.

\begin{table*}[t]
\centering
\begin{tabular}{p{3cm} p{2cm} p{2cm} p{2cm} p{2cm} p{2cm} p{2cm}}
\hline 
\hline 
 & \textbf{Q1} & \textbf{Q2} & \textbf{Q3} & $\textbf{Q1}^{*}$  & $\textbf{Q2}^{*}$  & $\textbf{Q3}^{*}$  \\ 
\hline 
Readout error& 0.40\% & 0.39\% & 0.46\% & 0.44\% & 0.50\%  & 0.48\% \\ 
Relaxation error& 0.46\% & 0.38\% & 0.40\% & 0.50\% & 0.42\% & 0.40\% \\
Separation error& 0.03\% & 0.017\% & 0.004\% & 0.17\% & 0.04\%  & 0.006\% \\ 
\hline
\end{tabular}
\caption{Readout error budget for the three qubits used in the experiments. Columns marked with asterisks (*) correspond to simultaneous three-qubit readout. The readout pulse length $\tau_{\mathrm{ro}}$, demodulation time $\tau_{\mathrm{demod}}$, and total measurement time $\tau_{m}$ are 100~ns, 200~ns, and 250~ns, respectively.}
\label{erroranalysis}
\end{table*}

\section{Readout leakage analysis}

\begin{figure*}[t]
\centering
\includegraphics[width=0.9\linewidth]{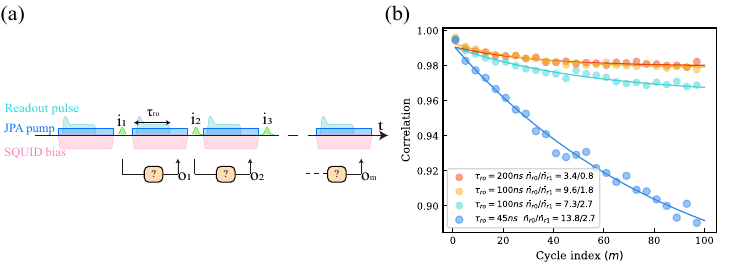}
\caption{Readout-induced leakage benchmarking using the random circuit method from Ref.~\cite{hazra2025benchmarking}. (a)~Pulse sequence for the readout-induced leakage benchmarking experiment. Readout operations are interleaved with randomized bit-flip operations.
(b)~Mean correlation as a function of the detection cycle index $m$ for Q2, using four different readout pulses. Data are averaged over 200 randomized sequences. Solid curves correspond to exponential fits.
}
\label{fig:leakage_analysis}
\end{figure*}

\begin{table*}[h]
\centering
\begin{tabular}{p{5cm} p{1.5cm} p{1.5cm} p{1.5cm} p{1.5cm} p{1.5cm} p{1.5cm} p{1.5cm} }
\hline 
\hline 
Readout Duration (ns) & 200 & 100 & 100 & 60  &  45 & 40 & 30 \\
$n_{r0}$/$n_{r1}$& 3.4/0.8 & 9.6/1.8 & 7.3/2.7 & 10.4/1.7  &  13.8/2.7 & 18.8/4.6 & 21.5/5.0 \\ 
\hline 
$\mathcal{L} = \mathcal{L}_\downarrow$+$\mathcal{L}_\uparrow$ (\%), This work &  2.20 & 1.78 & 1.37 & 0.79  &  0.94 & 2.64 & 6.40  \\ 
$\mathcal{L}_\downarrow$ (\%) &  2.12  & 1.70 & 1.27 & 0.64  &  0.51 & 0.37 & 0.49  \\ 
$\mathcal{L}_\uparrow$ (\%) &  0.08  & 0.08 & 0.10 & 0.15  &  0.43 & 2.27 & 5.91  \\ 
\hline
$\mathcal{L} = \mathcal{L}_\downarrow$+$\mathcal{L}_\uparrow$ (\%), Ref.~\cite{hazra2025benchmarking,marxer2025above}  &  3.08 & 3.07 & 1.82  &&  1.12 &   &   \\ 
$\mathcal{L}_\downarrow$ (\%) &  3.01  & 2.98 & 1.72 & & 0.78  &   &    \\ 
$\mathcal{L}_\uparrow$ (\%)  &  0.07  & 0.09 & 0.10 & & 0.34  &  &    \\ 
\hline
\hline
\end{tabular}
\caption{Readout-induced leakage rates for Q2 using different readout pulses. Each pulse is optimized for QND fidelity. Faster readout pulses with higher resonator photon numbers result in increased leakage rates $\mathcal{L}_\uparrow$ and decreased seepage rates $\mathcal{L}_\downarrow$.}
\label{QND leakage}
\end{table*}

We now derive Eq.~(3) in the main text. Let $P_m$ denote the population of the computational state before the $(m+1)$-th measurement, with the corresponding leakage population being $1 - P_m$. The population after the next measurement, $P_{m+1}$, is given by:
\begin{subequations}
\begin{align}
P_{m+1} &= P_{m} - P_{m} \mathcal{L}_\uparrow + (1 - P_m)\mathcal{L}_\downarrow \\
&= P_{m}(1 - \mathcal{L}_\uparrow - \mathcal{L}_\downarrow) + \mathcal{L}_\downarrow.
\end{align}
\end{subequations}
Here, $\mathcal{L}_\uparrow$ and $\mathcal{L}_\downarrow$ denote the leakage and seepage rates, respectively. Solving this recurrence relation yields an exponential decay for $P_m$ of the form:
\begin{equation}
P_m = A(1 - \mathcal{L}_\uparrow - \mathcal{L}_\downarrow)^m + B.
\label{eq:leakage_dynamics}
\end{equation}

We compare our readout-induced leakage benchmarking method with the random circuit method used in Ref.~\cite{hazra2025benchmarking,marxer2025above}. The pulse sequence for the randomized benchmarking is shown in Fig.~\ref{fig:leakage_analysis}(a). The method consists of $m$ repeated readout operations interleaved with randomized bit-flip operations. Bit-wise correlations are calculated by comparing the measurement bit-strings with the ideal ones determined by the bit-flip operations.
The mean correlation $\langle\mathcal{C}_m\rangle$ versus the cycle index $m$ for different readout pulses of Q2, averaged over different randomizations, are shown in Fig.~\ref{fig:leakage_analysis}(b).
Leakage and seepage rates during readout are extracted by fitting the model~\cite{marxer2025above}:
\begin{equation}
\langle\mathcal{C}_m\rangle=\frac{\mathcal{L}_\uparrow(A-\frac{1}{2})(1-\mathcal{L})^m+A\mathcal{L}_\downarrow+\frac{1}{2}\mathcal{L}_\uparrow}{\mathcal{L}}.
\end{equation}
Results for Q2 under different pulse conditions are summarized in Table~\ref{QND leakage}, along with those obtained by fitting Eq.~\ref{eq:leakage_dynamics}. The leakage rates extracted from both methods show good agreement. We attribute the minor observed discrepancies primarily to fitting errors.

\section{Scalable multiplexed readout with tunable broadband Purcell filters}

\begin{figure}[t]
\centering
\includegraphics[width=\linewidth]{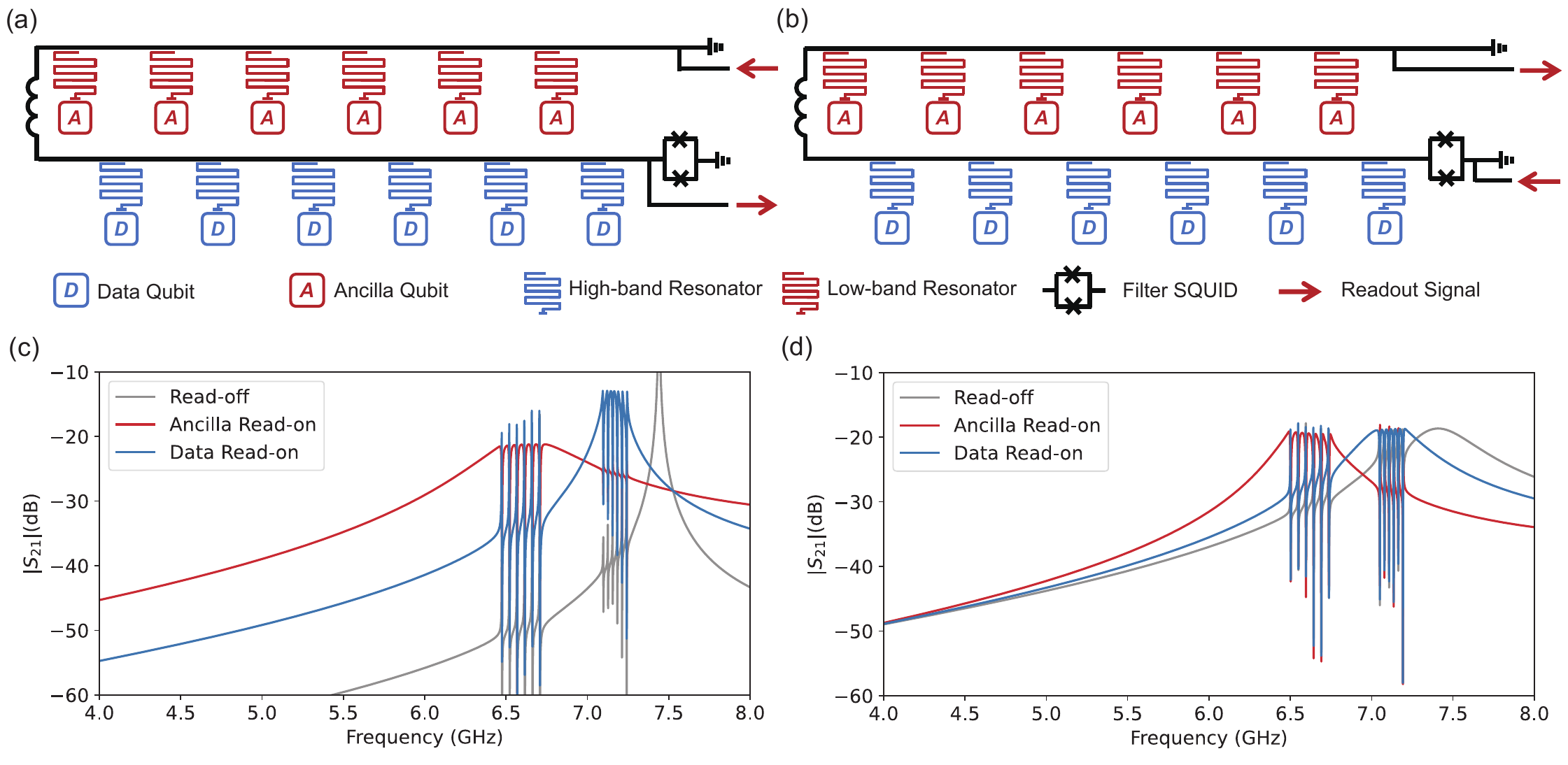}
\caption{
Multiplexed readout scheme based on a tunable broadband Purcell filter.
(a) Circuit diagram with a variable-bandwidth Purcell filter. Data qubits (blue squares) and ancilla qubits (red squares) share a common $\lambda/2$ Purcell filter, with resonators occupying distinct frequency bands. The filter bandwidth is made tunable by placing the SQUID between the readout port and ground.
(b) Circuit diagram with a fixed-bandwidth Purcell filter. The filter linewidth is fixed and determined by the electrical position of the readout port.
(c) Simulated transmission spectrum $|S_{21}|$ for three operational regimes: ancilla read-on (red), data read-on (blue), and read-off (grey). The average resonator linewidths for ancilla (data) qubits in these regimes are 15.3, 1.3, 0.1 (0.3, 4.7, 1.3)~MHz, respectively, corresponding to junction inductances of 0.65, 0.18, and 0.04~nH.
(d) Simulated transmission spectrum $|S_{21}|$ for the three operational regimes. The average resonator linewidths for ancilla (data) qubits are 13.4, 1.2, 0.6 (0.6, 6.3, 2.8)~MHz, corresponding to junction inductances of 0.48, 0.16, and 0.04~nH.
}
\label{multiplex_readout}
\end{figure}

We discuss the scalability of a multiplexed readout architecture based on tunable broadband Purcell filters within the context of quantum error correction (QEC). In conventional 2D surface code layouts, data and ancilla qubits typically employ separate feedlines to avoid parasitic measurement-induced dephasing~\cite{heinsoo2018rapid,lacroix2025scaling}.
By leveraging the tunable resonator linewidth enabled by the tunable Purcell filter, resonators for both data and ancilla qubits can be integrated on a single feedline while operating in distinct frequency bands, as illustrated in Fig.~\ref{multiplex_readout}(a) and (b) for both variable and fixed filter bandwidth configurations.
Dynamically tuning the filter frequency to the desired band enables selective measurement of either ancilla or data qubits.
During measurement, the resonator linewidth of the target qubits (ancilla or data) is optimized to match their dispersive shift, thereby maximizing the SNR. For unmeasured qubits, the filter is detuned from their resonator frequencies, resulting in a significantly reduced linewidth. This suppression of the resonator linewidth protects idle qubits from both thermal noise and parasitic measurement-induced dephasing~\cite{Gambetta2006,blais2021circuit,heinsoo2018rapid}.

We validate this multiplexed readout scheme using Simulation Program with Integrated Circuit Emphasis (SPICE), modeling the filter SQUID as a linear inductance~\cite{krantz2019quantum}. The simulated transmission response of the filter is shown in Fig.~\ref{multiplex_readout}(c–d). The filter operates in three distinct regimes:
(1) \textbf{Ancilla Read-on Regime}: the filter is tuned to resonance with the ancilla qubit resonators;
(2) \textbf{Data Read-on Regime}: he filter is tuned to resonance with the data qubit resonators;
(3) \textbf{Read-off Regime}: the filter is biased away from all resonators.
For the variable-bandwidth configuration, the simulated resonator linewidths for ancilla (data) qubits in the three regimes are 15.3, 1.3, 0.1 (0.3, 4.7, 1.3) MHz, corresponding to SQUID junction inductances of 0.65, 0.18, and 0.04 nH, respectively. 
This configuration is particularly suitable for surface code QEC, where ancilla qubits require frequent and rapid measurement for error detection, while data qubits are protected from ancilla readout tones due to their suppressed effective linewidth $\kappa_\mathrm{r}$~\cite{yan2018distinguishing}.
A similar trend is observed for the fixed-bandwidth case, where the average resonator linewidths for ancilla (data) qubits are 13.4, 1.2, 0.6 (0.6, 6.3, 2.8)~MHz, corresponding to junction inductances of 0.48, 0.16, and 0.04~nH.
We find that a higher ON/OFF ratio in resonator linewidth is achieved with variable-bandwidth configuration. Moreover, a narrower filter bandwidth in the read-off regime provides stronger $T_1$ protection for idle qubits. However, this protection is significantly reduced when the filter bandwidth is increased in the ancilla read-on regime, as also observed experimentally in Fig.~\ref{T1_protection}(b). This limitation can be mitigated by incorporating a band-stop filter to provide higher-order $T_1$ protection for the qubits~\cite{reed2010fast,zhou2024high}.


%